\documentclass[article,nofootinbib,twocolumn,showpacs]{revtex4}
\def\pl{{\mbox{\tiny{Pl}}}}
\def\st{{\mbox{\tiny{st}}}}
\def\dsr{{\mbox{\tiny{DSR1}}}}

\begin{document}

\title{Time delay of light signals in an energy-dependent spacetime metric }
\author{A.F. Grillo}
\affiliation{INFN - Laboratori Nazionali
del  Gran  Sasso,  SS.  17bis,  67010  Assergi  (L'Aquila)  -  Italy}
\author{E.  Luzio}
\affiliation{INFN - Laboratori Nazionali
del  Gran  Sasso,  SS.  17bis,  67010  Assergi  (L'Aquila)  -  Italy}
\author{F.  M\'endez}
\affiliation{Departamento de Fisica, Universidad de Santiago de
  Chile, Av. Ecuador 3493, Casilla 307 Stgo-2 - Chile}

\begin{abstract}
In  this  note  we  review  the  problem  of  time  delay  of  photons
propagating in  a spacetime with  a metric that explicitly  depends on
the energy of the  particles (Gravity-Rainbow approach).  We show that
corrections due  to this approach --  which is closely  related to DSR
proposal --  produce for small redshifts ($z<<1$)  smaller time delays
than in the generic Lorentz Invariance Violating case.
\end{abstract}
\maketitle

\section{Introduction}

The  idea that  relativistic symmetry  might not  be preserved  at all
energy scales has been a subject of an intense debate and study during
last  years.  Proposals  of how  to  modify the  Lorentz symmetry  and
models implementing this idea,  from which measurable consequences can
be obtained, have been investigated by large \cite{debate}.

In very general grounds, these  proposals can be divided in two types:
{\it  a)}  those  where  Lorentz  symmetry is  broken  by  choosing  a
preferred reference frame and {\it  b)} those were Lorentz symmetry is
deformed and  the relativistic principle is preserved.  In the present
note  we will  focus on  the consequences  for the  time of  flight of
photons in case (b).

Double Special  Relativity (DSR) \cite{dsr}  models fall in  case (b).
Generically,  they are non  linear realizations  of the  Lorentz group
that incorporate  a second invariant scale (momentum  or energy scale)
in order to solve the  following problem: if Lorentz symmetry is valid
only up to certain energy (or momentum) scale, then this scale must be
invariant for all observers on inertial reference frames.

Even  if this  idea  has concrete  realizations  for the  case of  one
particle in the momentum space, a consistent approach in spacetime and
multiparticle sector is still a matter of intense debate\cite{deb2}.

A  possible solution  to spacetime  problem is  the so  called Rainbow
Gravity \cite{ms,gm}.  Here, a spacetime is introduced that is dual to
the momentum space  where Lorentz group has a  non linear realization:
as  a  result,  the  metric  of this  spacetime  is  energy  dependent
\cite{mavro1}.  This approach admits  also curved spacetimes which are
solutions  of (modified)  Einstein Equations.  We will  refer  to this
space  (curved  and  energy  dependent)  as  {\it  rainbow  spacetime}
(\cite{ms}).

The  problem we  address here  is related  with this  modifications of
spacetime  structure and  the possibility  of testing  it  by redshift
and/or time of flight measurements \cite{mavro2}.  In concrete, we are
interested  in  the   modification  of  photon  redshiftsgenerated  by
DSR-like changes in the dispersion relation. 

Let us briefly review the standard case.  In the Cosmological Standard
Model,  the  metric  of  the   universe  is  given  by  the  Friedman-
Robertson-Walker line element
\begin{equation}
\label{FRW}
ds^2=dt^2-a(t)^2\left(\frac{dr^2}{1-k r^2}+d\Omega^2\right),
\end{equation}
where $t,r,\theta,\phi$ are the usual cosmological coordinates and $k$
is the three-dimensional space curvature ({which we will take equal to
zero from here on}).

Redshift $z$ is a wavelength (or frequency) shift due to the fact that
light  signals propagate  in  background (\ref{FRW})  and relates  the
wavelength of  the photon at emission ($\lambda$)  with the wavelength
$\lambda_0$ of the photon today (in cosmological terms),
namely
\begin{equation}
\label{rs}
z\equiv \frac{\lambda_0 -
\lambda}{\lambda}=\frac{\lambda_0 }{\lambda}-1
\end{equation}
or in terms of energy
\begin{equation}
\label{zenergy}
E=E_0(z+1).
\end{equation}
All these definitions are still  valid in the deformed case since they
do not depend on details  of propagation. The relation between $z$ and
the scale factor $a(t)$, however,  depends indeed of those details. In
concrete, from the fact that  metric has the shape (\ref{FRW}), -- see
for example  \cite{win}) -- since  the space is a  maximally symmetric
one  and  then  spatial  coordinates  can be  chosen  as  a  co-moving
reference system, one has
\begin{equation}
\label{ar}
\frac{a_0}{a}=\frac{\lambda_0}{\lambda}=z+1,
\end{equation}
with $a_0=a|_{z=0}$.
In the next section we will discuss how this property changes and explore the 
consequences for the calculation of proper distances and time delay of
photons.  

\section{Rainbow Redshift}
Consider now an energy dependent metric \footnote{Here we follow the notation 
of Magueijo and Smolin in \cite{ms}}
\begin{equation}
ds^2=-f^{-2} (E)~dt^2+g^{-2}(E)~a^2(t)~d\chi^2,
\end{equation}
with $d\chi^2=\gamma_{ij}dx^idx^j$ is the spatial line element. { Even
  if it is possible to  perform calculations for general functions $f$
  and $g$,  in the present  note we will  focus on the  propagation on
  this spacetime up  to first order in $M_\pl^{-1}$,  that is, we will
  take functions with the shape 
\begin{equation}
\label{fg}
f(E)\sim1+\phi~\frac{E}{M_\pl},~~~~~~~~~~
g(E)\sim1+\gamma~\frac{E}{M_\pl}
\end{equation}
where $\phi,\gamma$ are numerical constants of order 1. For example,
for a  DSR1 deformation (following the  classification in \cite{luki})
we have $\phi=0,\gamma=1/2$, 
while for DSR2 we have $f=g$  to all orders in $E/M_\pl$ and therefore
$\phi=\gamma$.}

Note  that the  definition  of redshift  (\ref{rs})  does not  change,
however its relation 
with $a$   (Eq.  (\ref{ar}) in the standard case)  does.  Indeed,
  if the relation between wavelength and momentum remains unchanged
  one has, considering the modified dispersion relation
$$
E^2f^2(E)-p^2g^2(E)=0=E^2f^2(E)-\frac{1}{\lambda^2(E)}g^2(E)
$$
and
$$
\frac{f(E_0)}{f(E)}~\frac{g(E)}{g(E_0)}~\frac{a(t_0)}{a(t)}=\frac{E}{
E_0}.
$$
The last equality is an assumption, depending on the (unknown)
QM in rainbow space-time. It is however natural, since it produces
wavelengths approaching the Plank length when energies approach the
Planck mass. 

Using the definition of $z$ we can write previous  expression as
\begin{equation}
\label{newa}
\frac{a(z)}{a_0}=\frac{f(E_0)}{g(E_0)}\frac{g(E_0(z+1))}{f(E_0(z+1))}
~\frac{1}{z+1}.
\end{equation}

To summarize, the definition of redshift in this rainbow spaces is
still (\ref{rs}), and this just says that to a measured energy $E_0$
corresponds an emitted energy $E$ at redshift $z$. Equation
(\ref{newa}), on the  other hand, says that to this redshift $z$ corresponds
a given value of $a$, which is determined by equations of motion.  

Notice that in this framework \cite{ms} photons of different
  energies see a different expansion.

In the following section we will use both relations to calculate the
proper distance traveled by a photon that is received at present,
with energy $E_0$. 

\subsection{Photons proper distances}
The comoving (proper) distance is the integral of equations of motion and for
simplicity we will consider only radial trajectories, that is
$$
r(z,E_0)=\int_{t}^{t_0}~\frac{g(E)}{f(E)}~\frac{dt'}{a(t')}.
$$
 Here, the speed of photons is energy dependent through  $f$ and $g$;
 the constant factor $c$ has  been chosen unity. The previous equation
 can be rewritten as an integral  in $z$ as in the standard case. From
 (\ref{newa}) we have 
\begin{eqnarray}
r(z,E_0)&=&\frac{g_0}{f_0a_0}\int_0^z \frac{g(E_0(z'+1))}{f(E_0(z'+1))} 
\nonumber 
\\
&&\times
\frac{d}{dz'}\bigg[(z'+1)\frac{f(E_0(z'+1))}{g(E_0(z'+1))}\bigg]\frac{dz'}{H 
(z')},\nonumber
\end{eqnarray}
where  $f_0=f(E_0),~g_0=g(E_0),~a_0=a(t_0)$ and $H(z)=\dot{a}(t)/a(t)$
and it is 
given by \cite{ms} 
\begin{equation}
\label{H}
H=\left[\frac{8\pi}{3}G(E)\frac{\rho}{f^2}                            +
  \frac{\Lambda(E)}{3}\right]^{\frac{1}{2}}.  
\end{equation}
$G$ and $\Lambda$ can be, in principle, functions of the energy.

It is possible  to rewrite $r(z,E_0)$ in order  to show explicitly the
modifications due  to the energy  dependence of the metric.   A direct
calculation allows us to write
\begin{equation}
\label{finr}
r(z,E_0) =\frac{g_0}{f_0a_0}\int_0^z \bigg[1-(z'+1)\frac{d}{dz'}
\ln\left(\frac{g}{f}\right)\bigg]\frac{dz'}{H(z')},
\end{equation}
where $f, g$ and $H$ are evaluated in $E_0(z'+1)$.

{ This last expression is  valid for any symmetry deformation, however
in  this general  form  is  not useful  to  extract information  about
possible  physical consequences.  Since  we are  interested in  linear
corrections,   we   can  circumvent   this   problem  by   considering
deformations of the type (\ref{fg}), namely
\begin{equation}
\label{GL}
G(E)\sim G\left(1+\Gamma\frac{E}{M_\pl} \right),~~~
\Lambda(E)\sim \Lambda\left(1+\lambda\frac{E}{M_\pl} \right),
\end{equation}
with $G,\Lambda$  the Newton and cosmological constants  (in the limit
$M_\pl\to\infty$) and  $\Gamma,\lambda$, numerical constants  of order
1. 

With (\ref{fg}) and (\ref{GL}) we can calculate explicitly first order
corrections to proper distances in (\ref{finr}), namely 
\begin{eqnarray}
\label{rone}
&&r(z,E_0)=\frac{1}{a_0}\int_0^z\frac{dz'}{H_\st(z')}-\nonumber
\\
&&-\frac{E_0}{M_\pl
  a_0}\int_0^z\frac{dz'}{H_\st(z')}\left(z'(\gamma-\phi)+(z'+1) 
\frac{\delta H^2(z')}{2H_{\st}^2(z')}\right)\nonumber,
\\
&&
\end{eqnarray}
where   $H_\st$   is   the   standard  Hubble   parameter   (that   is
$f=1,G(E)=G,\Lambda(E)=\Lambda$  in (\ref{H}))  while $\delta  H^2$ is
the first order correction due to the dependences on $E$ 
\begin{equation}
\delta H^2=\frac{8\pi}{3}G\rho(\Gamma-2\phi)+\frac{\Lambda~\lambda}{3}.
\end{equation}

Let us consider the example of DSR1. We have 
\begin{eqnarray}
\label{rdsr1}
&&r_\dsr (z,E_0)=\frac{1}{a_0}\int_0^z\frac{dz'}{H_\st(z')}-\nonumber
\\
&&-\frac{E_0}{2M_\pl a_0}\int_0^z\frac{dz'}{H_\st(z')}\left(z'+(z'+1)
\frac{\delta H^2(z')}{H_{\st}^2(z')}\right). \nonumber
\\
&&
\end{eqnarray}
Note  that  the  behavior   of  this  function  depends  on  functions
$G,\Lambda$. In  fact, if we  consider (a very  conservative approach)
$\Gamma=0=\lambda$, we have 
\begin{equation}
\label{finrdsr}
r_{\mbox{\tiny{DSR1}}}(z,E_0)=\frac{1}{a_0}\int_0^z
\bigg[1-z'\frac{E_0}{2M_\pl} 
\bigg]\frac{dz'}{H_\st(z')}.
\end{equation}
which  is different  from one  obtained  recently by  Jacob and  Piran
\cite{jp} and  the reason, in this  particular case, is  that they use
the standard relation between $a$ and $z$ while for us, it is given by
(\ref{newa}).   In our case,  this introduces  an extra  factor $g_0$,
which cancels the 1 in $z'+1$.  For this particular example our result
coincides with that in \cite{jp} only for $z>>1$.

However,  a  rather different  behavior  appears  if  we consider  non
vanishing $\Gamma$ and $\lambda$. In fact, since they are of the order
1,  then the  last term  in  (\ref{rdsr1}) is  of order  one, that  is
$\delta H^2(z')/2H_{\st}^(z')\sim 1$ and  then a correction similar to
the one  obtained in  \cite{jp} is obtained.  An illustrative  case is
$\Gamma=\Lambda\equiv \sigma$ and it gives
\begin{equation}
\label{finrdsr1}
r_{\mbox{\tiny{DSR1}}}(z,E_0)=\frac{1}{a_0}\int_0^z
\bigg[1-\frac{E_0}{2M_\pl}(z'+(z'+1)\sigma) 
\bigg]\frac{dz'}{H_\st(z')}.
\end{equation}
We will return on that in the discussion  section. }

{For  DSR2, instead,  the only  possible corrections,  in  the present
  approach, come from functions $G(E),\Lambda(E)$. In fact, we have in
  (\ref{rone}) 
\begin{eqnarray}
\label{rdsr2}
r(z,E_0)&=&\frac{1}{a_0}\int_0^z\frac{dz'}{H_\st(z')}-\nonumber
\\
&&-\frac{E_0}{M_\pl a_0}\int_0^z\frac{dz'}{H_\st(z')}(z'+1)
\frac{\delta H^2(z')}{2H_{\st}^2(z')}.
\end{eqnarray}
and for  $\Gamma=0=\lambda$ no corrections are  obtained. Instead, for
the other case discussed  before $\Gamma=\lambda\equiv \sigma$ we have
a correction of the type obtained in \cite{jp}. 

}

\subsection{Photons time delay}
The time of flight of a photon that travels between two points labeled
by $t$ and $t_0$ is 
$$
\delta t= \int_{t_0}^t dt'=\int_{0}^z \frac{da}{H(z')a},
$$
were we have chosen $t_0$ as the present time. Using (\ref{newa}) the
lookback time is:
 \begin{equation}
 \label{dt}
 \delta                                                t=\int_0^z\bigg[
 -\frac{1}{z+1}+\frac{d}{dz'}\ln\left(\frac{g}{f}\right) 
 \bigg]\frac{dz'}{H(z')},
 \end{equation}
where $f$ and $g$ are  evaluated in $E_0(z'+1)$. This expression gives
the time that a photon takes to travel from a source at a given $z$ to
the present  (with $z=0 $) if  the energy measured now  is $E_0$ (what
means of course that the energy at the emission time were $E_0(z+1)$).

Consider  now two  photons  produced at  $z$  with different  energies
there,  arriving at present  time ($z=0$)  (of course,  with different
energies). In  the standard case,  only the first  term in the  RHS of
(\ref{dt}) is present and it  does not depend on energy, therefore the
difference on time  of flight between these two  photons will be zero,
as is well known. However, in  the present case, due to the dependence
on the  final energy (second term  in RHS of (\ref{dt}))  we will have
the following difference for the lookback time

\begin{equation}
\label{delt}
\Delta                t                =\int_0                ^z\Delta
 \bigg[\frac{1}{H(z')}\frac{d}{dz'}\ln\left(\frac{g}{f}\right) 
 \bigg] dz',
\end{equation}
with 
\begin{eqnarray}
\Delta \bigg[\frac{1}{H(z')}\frac{d}{dz'}\ln\left(\frac{g}{f}\right)  \bigg]
&\equiv&
\frac{1}{H^{(1)}}\frac{d}{dz'}                                \ln\left(
\frac{g(E_0^{(1)}(z'+1))}{f(E_0^{(1)}(z'+1))}\right)\nonumber 
\\
& -& 
 \frac{1}{H^{(2)}}\frac{d}{dz'}\ln                               \left(
 \frac{g(E_0^{(2)}(z'+1))}{f(E_0^{(2)}(z'+1))} \right),\nonumber 
\end{eqnarray}
where $E_0^{(1)}$ and $E_0^{(2)}$ are the energies of the two photons, 
measured at $z=0$ and $H^{(i)}$  is $H$ defined in (\ref{H}) evaluated
in $E_0^{(i)}$, for $i=1,2$. 

{The   previous   expression    is   valid   for   general   functions
$f,g,G,\Lambda$.  In order  to  analyze  the behavior  of  it we  will
consider again  only first order contributions,  namely (\ref{fg}) and
(\ref{GL}). Note  that, since the  derivative of $\ln(g/f)$ is  of the
order $E/M_\pl$, then contributions due to $\Gamma$ and $\lambda$ will
not be present.  In fact, a straightforward calculation shows that, in
the linear approach, (\ref{delt}) becomes
$$
\Delta t =\frac{\gamma-\phi}{M_\pl}\int_0 ^z\Delta \bigg[\frac{1}{H_\st(z')}E_0
 \bigg] dz',
$$
but $H_\st$ does not depend on the energy of particles, then
\begin{equation}
\label{delt1}
\Delta t =\frac{\Delta E_0(\gamma-\phi)}{M_\pl}\int_0 ^z \frac{dz'}{H_\st(z')},
\end{equation}
}
We see again that, for DSR2  there will be no (energy dependent) delay
while for DSR1 we have 
\begin{equation}
\label{dsrdt}
\Delta t = \frac{\Delta E_0}{2M_\pl} \int_0 ^z\frac{dz'}{H(z')},
\end{equation}
which coincides  with the result of \cite{ellis}.  {Corrections due to
  the dependence  of $G$ and  $\Lambda$ on $E$  turn out to  be second
  order in $E/M_{pl}$}. 

A similar question can be formulated about differences on proper
distances. Namely, two  photons produced by a source  at $z$, but with
different energies  there -- and therefore with  different energies at
$z=0$ -- will they have a  shift on their proper distances? The answer
for this rainbow spacetime can be obtained directly from (\ref{finr}),
but it is  not so illuminating and therefore we  will consider again {
  the linear approach, that is (\ref{rone}).  It is straightforward to
  see that in this case 
\begin{eqnarray}
\label{delrone}
\Delta           r(z,E_0)&=&          \frac{\Delta          E_0}{M_\pl
  a_0}\int_0^z\frac{dz'}{H_\st(z')}\bigg(z'(\gamma-\phi) 
\nonumber
\\
&&+(z'+1) \frac{\delta H^2(z')}{2H_{\st}^2(z')}\bigg)\nonumber,
\end{eqnarray}
Clearly  we  can define  a  time  delay  $\bar{{\Delta}} t=a_0  \Delta
r(z,E_0)$, and then 
\begin{eqnarray}
\label{deltone}
\bar{{\Delta}}                    t&=&                    \frac{\Delta
  E_0}{M_\pl}\int_0^z\frac{dz'}{H_\st(z')}\bigg(z'(\gamma-\phi) 
\nonumber
\\
&&+(z'+1) \frac{\delta H^2(z')}{2H_{\st}^2(z')}\bigg).
\end{eqnarray}

}
Some comments are in order here. First, the fact that a time delay can
be defined {proportional  to} the proper distance depends  also on the
relation between  $a$ and $z$, which  in our case is  not the standard
one; however this gives rise to second order corrections in $1/M_\pl$,
which have  been discarded in  the examples considered in  the present
note.  Second,  we would like to  call the attention on  the fact that
$\Delta t$  in (\ref{delt1})  and $\bar{\Delta} t$  in (\ref{deltone})
are different because they measure different physical properties.  The
first is related  to actual measurements of times  while the second is
based on measurements of proper distances.

To finalize the present discussion, let us  point out that our
results are strongly dependent on the choice of $G(E),\Lambda(E)$. 

\section{Discussion and conclusions}

In this note we have explored the consequences on the determination of
(energy  dependent)  proper distances  and  arrival  times of  photons
produced  by sources  at  redshift $z$  in  Doubly Special  Relativity
models with modified dispersion relations.

The major  problem to  perform this kind  of calculation is  the still
uncertain knowledge  of the  spacetime structure compatible  with this
symmetry deformation.  As we pointed  out in the introduction, this is
an open problem, even if much work (and progress) has been done.

On  the  other hand  there  exist  (partial)  proposals (for  instance
\cite{ms}) that allow  to treat at least in a  self consistent way the
problem of the propagation of particles in cosmological space-times.

For the discussion  carried out in the present note,  we have used the
proposal of Magueijo and Smolin  where the spacetime metric depends on
the energy of  the particle that probes this  spacetime (the so called
\lq rainbow metric\rq) and satisfies (modified) Einstein equations for
some matter distribution. In this  approach photons move on a modified
geodesics  as a  consequence  of the  modification  of the  dispersion
relation, and experience a  modified and energy dependent cosmological
expansion.

In this  context we find that  (at least for photons  emitted at small
$z$) the effects  of the modifications on the  photon geodesics and of
the scale factor  do compensate and one obtains for  the time delay of
photons of different energies emitted  at same $z$ the result reported
by  Ellis et  al.  in \cite{ellis},  in  the particular  case of  DSR1
photons propagating in a (deformed) FRW universe with constant $G$ and
$\Lambda$.

As was pointed out in previous section, our results depend strongly on
the  funtional form of  $G(E)$ and  $\Lambda(E)$ (as  well as  that of
$f(E)$ and  $g(E)$); this just  reflects the strong dependence  of the
time delay on the structure of spacetime.

{In concrete, we have shown that, in this model, for $G$ and $\Lambda$
  constant,  the  DSR1 approach  and  ref.  \cite{jp}, give  different
  results  for small  $z$, while  they  coincide for  large values  of
  redshift (see relation (\ref{newa})). On the other hand, with linear
  corrections to  $G$ and $\Lambda$, the  modifications can compensate
  and  then a  similar  result  to \cite{jp}  is  obtained. For  DSR2,
  instead, possible corrections arise only from this last effect.

For the lookback time,  corrections from the dependence of $G,\lambda$
on the energy  of photons, are second order  effects in $M_{\pl}^{-1}$
and then  only relation (\ref{newa})  is relevant. We have  shown that
time  delays   calculated  in  this  way  coincide   with  results  in
\cite{ellis}. }

Finally, we would like to  emphasize that our result is different from
that  recently  reported  by  Jacob  and Piran  \cite{jp},  but  there
different  hypotheses on the  space-time structure  are made.  This in
principle leaves open the possibility of experimentally distinguishing
among different phenomenological consequences of Quantum Gravity.

\acknowledgments

One of  us (F.M)  would like  to thanks L.N.G.S.  (INFN) for  the kind
hospitality  during the  preparation of  this manuscript  and  for the
grant Fondecyt-Chile 1060079.

\end{document}